\documentstyle[aps,prl,twocolumn]{revtex}
\include{psfig}

\begin{document}
\draft

\title {Configuration space connectivity across the
fragile to strong transition in silica}

\author{Emilia La Nave$^1$, H. Eugene Stanley$^{1}$ and 
Francesco Sciortino$^{2}$}

\address{$^1$Center for Polymer Studies, Center for Computational
Science, and Department of Physics, Boston University, Boston, MA
02215 USA}

\address{$^2$Dipartmento di Fisica e Istituto Nazionale per la Fisica
della Materia, Universit\'{a} di Roma ``La Sapienza'', Piazzale Aldo
Moro 2, I-00185, Roma, Italy}

\date{\today}

\maketitle
\begin{abstract}

We present a numerical analysis for SiO$_2$  
of the fraction of diffusive direction $f_{\mbox{\scriptsize diff}}$ for temperatures  $T$
on both sides of the fragile-to-strong crossover.
 The  $T$-dependence of $f_{\mbox{\scriptsize diff}}$ clearly reveals this change in
 dynamical behavior. We find that for $T$  above the crossover
(fragile region) the system is always close to ridges  of the
potential energy surface (PES), while below the crossover (strong region),
the system mostly explores the PES local minima. 
Despite this difference, the power law  dependence of $f_{\mbox{\scriptsize diff}}$ on  the diffusion
constant, as well as the power law dependence  of $f_{\mbox{\scriptsize diff}}$ on 
 the configurational entropy, shows no change at the fragile to strong 
crossover.

\end{abstract}
\bigskip

A useful approach for relating dynamics and thermodynamics of
glassforming liquids in their supercooled states is offered by the study
of the liquid's potential energy surface (PES)\cite{ds2001}.  The PES is
the multidimensional surface generated by the system's potential energy
as a function of all atomic coordinates.  Numerical and theoretical
studies
\cite{goldstein,stillingerweber,sastry,sciortino,Kob00,keyes-pre,heuer,harrowell,jund}
are providing evidence that several aspects of the liquid dynamics have
a clear signature in the properties of the explored regions of the
PES.  For example, correlation functions
display stretching in time in the same temperature ($T$) range in which the
system explores regions of the PES associated with local minima
of deeper and deeper energy\cite{sri}.  Similarly, the rapid
slowing down of the dynamics which takes place in the weakly supercooled
region is associated to the exploration of regions of the PES close to
saddles of lower and lower
order\cite{keyes,st,lanave,angelani,cavagna,wales}.  
Recent numerical studies
for fragile liquids\cite{s-f} 
have suggested the possibility that 
a cross over $T$, $T_\times$
marks a change in the geometrical properties of 
the PES explored.
In this picture, above $T_{\times}$ the system
trajectories are mostly located close to the ridges between different
PES basins (``border dynamics''), while below $T_{\times}$ the system only
rarely samples the regions connecting different PES basins
(``minimum-to-minimum dynamics''). 
$T_\times$ has been associated\cite{angellmct,sokolov,st,lanave} with the critical
temperature predicted by the ideal mode-coupling theory (MCT)\cite{mct}.

Instantaneous normal mode (INM) analysis \cite{keyes} is a powerful
technique to investigate the connectivity properties of the PES explored
at different temperatures.  The curvature of the PES
along each of the $3N-3$ independent directions (the eigenvectors of the
Hessian matrix) is calculated and analyzed to estimate the fraction
of diffusive directions $f_{\mbox{\scriptsize diff}}$.  If the
representative configuration is crossing a ridge separating different
PES basins, some of the local curvatures are negative.  
A computationally-expensive but well-defined screening procedure 
to sort out the negative-curvature
directions contributing to diffusion\cite{keyes-pre} 
has been developed\cite{berne,claudio,lanave}. 
For all model potentials for
which such analysis has been performed, it has been shown that the $T$
and density $\rho$ variation of $f_{\mbox{\scriptsize diff}}$ controls
the behavior of the long time dynamics, supporting the hypothesis that
information about the local properties of the PES may be sufficient to
describe long-time dynamical processes.

The analysis of the PES connectivity  has been limited to models
for fragile glass-forming liquids and to temperatures above $T_{\times}$, due to
the difficulty in generating equilibrium configurations at low $T$.  In
all studied cases, it was consistently found that $T_{\times}$ locates the
$T$ at which $f_{\mbox{\scriptsize diff}}$ appears to extrapolate to
zero\cite{st,claudio,lanave}.  

Here, we present 
an evaluation of the fraction of
diffusive directions (using the INM analysis) for
the well-studied BKS model for silica\cite{bks}, for which we generate
{\bf equilibrium} configurations for temperature both above and below
$T_{\times}$. The INM spectrum of BKS silica has been previously 
calculated by Benbenek and Laird\cite{inmbks} for
equilibrium states above $T_{\times}$ but no evaluation of $f_{\mbox{\scriptsize diff}}$
was reported.
For the BKS model, the high $T$ dynamics has been shown to be
consistent with the predictions of MCT at both a qualitative\cite{Horbach99}
 and quantitative\cite{sk} level. At lower $T$,
the $T$-dependence of the characteristic times\cite{Horbach99}
shows a crossover toward an Arrhenius $T$ dependence  which has
been interpreted as a clear case of fragile-to-strong
transition\cite{peterbks}. 
The crossover temperature $T_\times$, around 3330 K at $\rho=2.36$~g/cm$^3$,
has been interpreted by
Horbach {\it et al} at as the MCT critical temperature\cite{Horbach99}.
In the Arrhenius region, the activation energy is  $54000 K$ and $60000 K$ , for 
oxygen and silicon respectively.

We find that  both the $T$ dependence of the diffusion
constant $D$ and of $f_{\mbox{\scriptsize diff}}$ show a clear
signature of the two different dynamical behaviors --- fragile above and strong below $T_{\times}$ ---.
 Despite these differences, the relation between $D$
and $f_{\mbox{\scriptsize diff}}$ is {\it not} sensitive to the presence
of dynamical changes. Moreover, we find that $f_{\mbox{\scriptsize
diff}}$ does not vanish around $T_{\times}$ but changes to an
Arrhenius $T$ dependence, similar to the $T$ dependence of  $D$.
 Finally, we show that the number of
diffusive directions is related to the number of basins of the
PES, providing a possible explanation for the recently-observed
validity\cite{sastry,peterbks,waternature} of the Adam-Gibbs
equation\cite{ag} in the $T$ region where border dynamics dominates.

Our results are based on extensive simulations of a system of 999 atoms, for 
$\rho=2.36$~g/cm$^3$, close to the density of ambient
pressure silica. We investigate eleven $T$, in the range from $T=2650 K$ to
$T=7000 K$.  We carry out simulations in the NVE ensemble, using 
a 1 fs integration timestep. We
evaluate the long range interaction by implementing the Ewald summation.
 To guarantee proper equilibrium conditions all
simulations lasted longer than several times the slowest collective
structural relaxation time; low $T$ runs lasted longer than 50 ns. We also
performed averages over eight different realizations.
For each studied state point we calculate 
eigenvalues and eigenvectors for  96 configurations. 
For each of the eleven $T$ studied,  we performed approximately
2600 minimizations, using a total of about 50,000 CPU hours.

\begin{figure}[htbp]
\begin{center}
\mbox{\psfig{figure=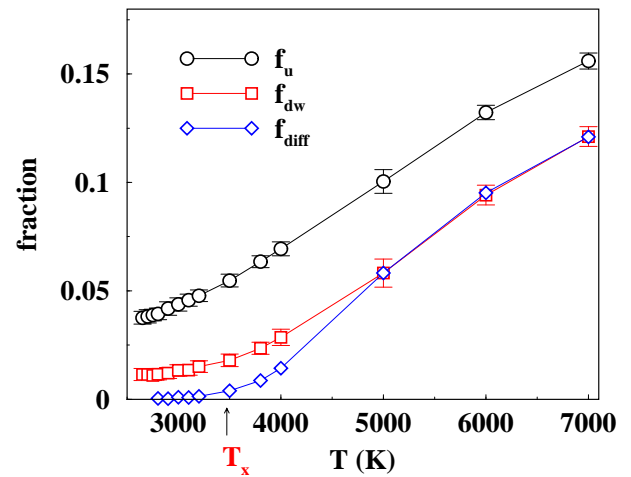,width=8cm,angle=-90}}
\end{center}
\caption{Temperature dependence for the
BKS model  of the 
fraction of modes with negative curvature $f_{u}$, 
the fraction of modes with double-well shaped 
one-dimensional profile $f_{\mbox{\scriptsize dw}}$, and the fraction  of 
diffusive directions $f_{\mbox{\scriptsize diff}}$. 
The arrow marks $T_{\times}$  for the isochore studied\protect\cite{Horbach99}.
}
\label{fig:frazioni}
\end{figure}

The INM analysis requires the evaluation of the eigenvectors and
associated eigenvalues of the potential energy second derivative matrix,
 the Hessian.  According to the procedure described in
Refs.\cite{berne,lanavelong}, all eigenvectors associated with negative
eigenvalues are inspected in order to eliminate those associated with
intra-basin anharmonicities (``shoulder'' and ``false barrier'' modes).
Fig.~\ref{fig:frazioni} shows (i) the $T$ dependence of the fraction of
directions with negative eigenmodes $f_{u}$,(ii) the fraction of directions
whose one-dimensional profile is double-well shaped $f_{\mbox{\scriptsize dw}}$, and (iii) the
number of diffusive directions $f_{\mbox{\scriptsize diff}}$,
calculated by eliminating from the double-well set all false barrier
modes\cite{berne,spiegazioni}.  Above $T=3330 K$, all three quantities
show a fast decrease with $T$.  While $f_u$ and $f_{\mbox{\scriptsize dw}}$  
assume non-zero values, $f_{\mbox{\scriptsize diff}}$
appears to approach zero on cooling. A clear change of concavity in the
$T$ dependence of $f_{\mbox{\scriptsize diff}}$ takes place  above
$T=3330 K$. As found previously for fragile liquids for $T>T_{\times}$, the
fast decrease of  $f_{\mbox{\scriptsize diff}}$ 
confirms that above $T=3330 K$: (i)
the slowing down of dynamics is associated with
 a progressive decrease of the number of possible directions that
lead to a different basin; (ii) the system trajectories are
located close to PES ridges ({\it border dynamics});
(iii) even for BKS silica, the dynamics properly described by
the ideal MCT are border dynamics, supporting the identification
of $T_\times$ with the MCT critical temperature.

\begin{figure}[htbp]
\begin{center}
\mbox{\psfig{figure=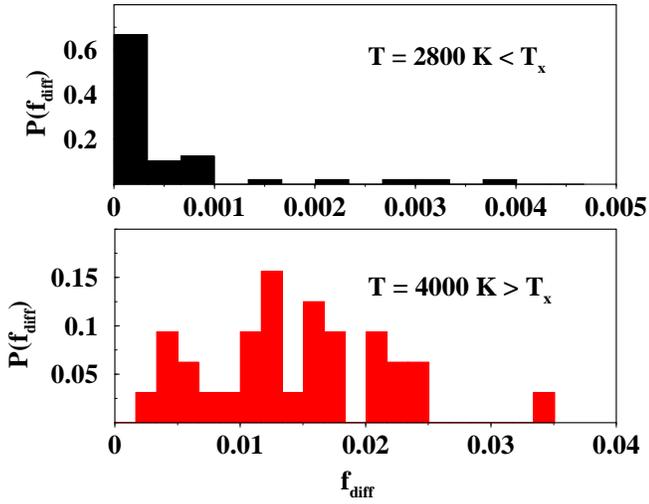,width=8cm,angle=-90}}
\end{center}
\caption{ 
Probability $P(f_{\mbox{\scriptsize diff}})$ of
finding a  configuration 
with a given $f_{\mbox{\scriptsize diff}}$ value 
above $T_{\times}$  ($T=4000 K$) 
and below $T_{\times}$ ($T=2800 K$).
While for $T>T_{\times}$ all examined configurations have a
nonzero $f_{\mbox{\scriptsize diff}}$, below $T_{\times}$, the distribution is
peaked around the origin.}
\label{fig:hysto}
\end{figure}

For $T$ below $3330 K$, the system spends
most of the time far from the PES ridges. To support this
statement, Fig.~\ref{fig:hysto} shows the probability $P(f_{\mbox{\scriptsize diff}})$ of
finding a 
configurations with a specific $f_{\mbox{\scriptsize diff}}$ value for
$T=2800 ~K$ ($<T_{\times}$) and $T=4000~K$ ($>T_{\times}$).  
Above  $T_{\times}$ all examined
configurations have a nonzero $f_{\mbox{\scriptsize diff}}$, while below
$T_{\times}$ the distribution is peaked around zero and most of
the  configurations are characterized by the absence of
diffusive directions.

\begin{figure}[htbp]
\begin{center}
\mbox{\psfig{figure=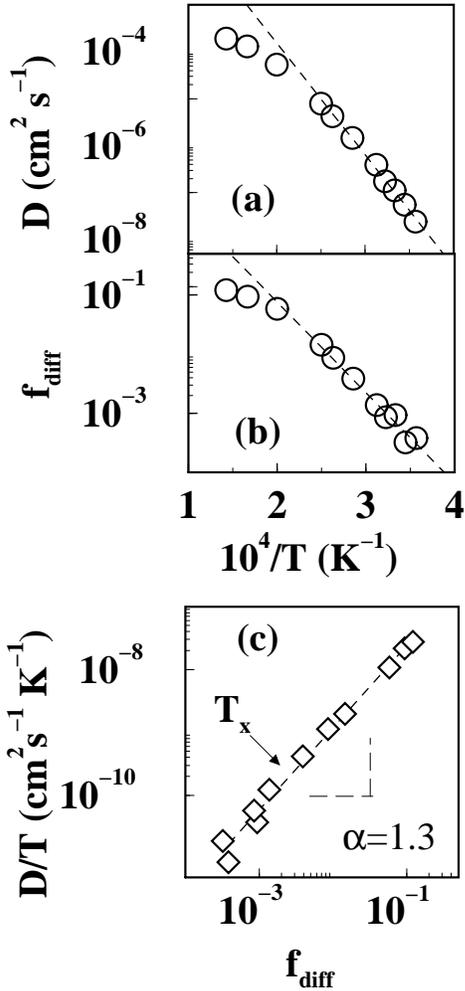,width=6cm,angle=-90}}
\end{center}
\caption{Arrhenius plot of (a) the diffusion constant $D$ for Si atoms; 
(b)  $f_{\mbox{\scriptsize diff}}$. 
The line has slope $54,000 K$  in (a) and $36,000 K$
 in (b) . 
Part (c) shows the parametric relation
$D/T$ vs $f_{\mbox{\scriptsize diff}}$ in a log-log scale. The data are
 perfectly smooth through the transition at $T_{\times}$.}
\label{fig:arrhe}
\end{figure}

We next investigate the functional relation between
$f_{\mbox{\scriptsize diff}}$ and  $D$.  
The $T$ dependence of both quantities,  shown in an Arrhenius plot in
Figs.~\ref{fig:arrhe}(a) and~\ref{fig:arrhe}(b), show a change
in the $T$-dependence above and below $T_{\times}$.
Despite this different dynamical behavior, the quantity
$f_{\mbox{\scriptsize diff}}$ is related to
$D$ by the same power law relation both for $T>T_{\times}$ and for $T<T_{\times}$ (Fig.\ref{fig:arrhe}(c)).
 In the entire studied $T$ region, $D$  follows the law 
\begin{equation}
D/T \sim (f_{\mbox{\scriptsize diff}})^\alpha ,
\end{equation}
 with $\alpha \approx 1.3 \pm 0.2$ over more than two
decades in $f_{\mbox{\scriptsize diff}}$ and more then three decades in $D/T$.
 The same
functional form describes the relationship between $D$ and
$f_{\mbox{\scriptsize diff}}$ both above and below $T_{\times}$ 
showing that
while the $T$ dependence of both $D$ and $f_{\mbox{\scriptsize diff}}$
is sensitive to the microscopic mechanisms controlling the dynamics, the
fragile-to-strong transition does not affect in the relation between  $D$ and
$f_{\mbox{\scriptsize diff}}$. We stress also that the same
functional form has been found to describe the
relation between $D$ and $f_{\mbox{\scriptsize diff}}$ in the case of the extended simple point charge
 (SPC/E) model for water\cite{lanave},
for which calculations were limited to $T>T_{\times}$.

\begin{figure}[htbp]
\begin{center}
\mbox{\psfig{figure=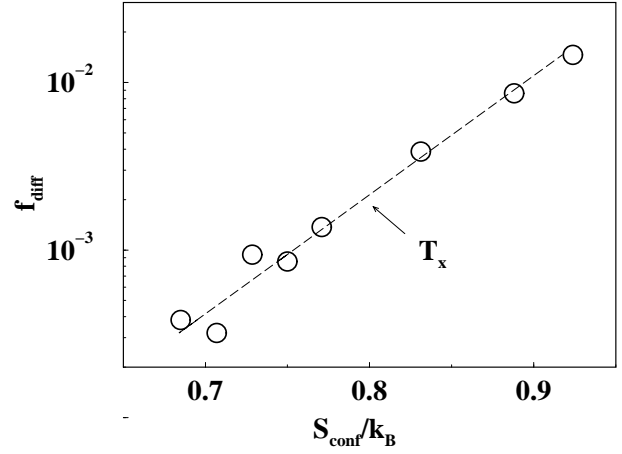,width=8cm,angle=-90}}
\end{center}
\caption{Parametric plot of the fraction of diffusive modes $f_{\mbox{\scriptsize diff}}$ as a function of the configurational entropy $S_{\mbox{\scriptsize conf}}$ for the BKS model of silica}
\label{fig:entro}
\end{figure}

The interpretation of the dynamics in terms of the fraction of diffusive
directions is complementary to the analysis which attributes the slowing
down of the dynamics  to the decrease of the liquid
configurational entropy $S_{\mbox{\scriptsize conf}}$\cite{ag}.
  In the case of BKS silica, $S_{\mbox{\scriptsize conf}}$ has 
been recently calculated and shown
to describe, via the Adam-Gibbs relation, the slowing down of the
dynamics both above and below $T_{\times}$ \cite{peterbks}.  Since
$S_{\mbox{\scriptsize conf}}$ is a measure of the number of distinct PES
basins explored by the system --- if  the description in terms of
$S_{\mbox{\scriptsize conf}}$ and the description in terms of
$f_{\mbox{\scriptsize diff}}$ are both valid --- a relation must exist
between the number of basins and the number of directions connecting
them.  Fig.\ref{fig:entro} shows the relation between
$S_{\mbox{\scriptsize conf}}$ and $f_{\mbox{\scriptsize diff}}$ for the
case of BKS silica.  Within the numerical uncertainty, the fraction of
diffusive directions appears to be proportional to the number of 
explored basins,
$\Omega \equiv e^{S{\mbox{\scriptsize conf}}/k_B}$, 
irrespective of the strong or fragile character of the dynamics.
Data in Fig.\ref{fig:entro} are consistent with similar findings
which were limited to the SPC/E model for water
to the $T$-region above $T_{\times}$.  The present results suggest that the
linear relation between $\log f_{\mbox{\scriptsize diff}}$ and
$S_{\mbox{\scriptsize conf}}$ is not model dependent; indeed it has been
recently derived within the random energy model \cite{remkeyes}.  The fact that
this relation holds even above $T_{\times}$ is particularly
interesting; since  above $T_{\times}$ the system
dynamics is a dynamics of borders, there is no clear reason why such
border dynamics should be well described by the Adam-Gibbs relation which
focuses on the number of basins explored as a function of $T$.  The
observed relation between $\log f_{\mbox{\scriptsize diff}}$ and
$S_{\mbox{\scriptsize conf}}$ may offer a key for the resolution of
this apparent paradox.  It is a challenge for future studies to find out if
$f_{\mbox{\scriptsize diff}}$ is a potentially richer quantity for
describing dynamics in deep supercooled states, as the results 
reported here seem to suggest.

We thank W. Kob, T. Keyes, G. Ruocco, S. Sastry, and A. Scala for
helpful discussions, and NSF Grant CHE-0096892 for support. FS
acknowledges support from MURST (PRIN 2000) and INFM (PRA-HOP and
Initiative Parallel Computing).

\begin{references}
\bibitem{spiegazioni}
 To separate the double-well directions into diffusive
 modes (modes connecting distinct minima, or inherent
structures), and false modes (intra-basin modes) we perform a conjugate
gradient minimization with tolerance $10^{-11} kJ/mol$, starting from the two
one dimensional double-well minima.  We then calculate the distance between the
two inherent structures $d \equiv [\sum_i^N (r^0_i-r^1_i)^2/N]^{1/2}$ ,
where $r^\alpha_i$ is the $i^{th}$ atom position in inherent structure
$\alpha$ .  The distribution of $d$ found is
bimodal, and the two contributions are 
separated by more than one order of magnitude.
Modes which contribute to the peak located at ``large'' $d \approx
0.01$ ~nm are classified as diffusive modes ($f_{\mbox{\scriptsize diff}}$).  All
other modes are classified as
``false'' double-well modes, since both sides of the one dimensional profile
end up in the same minimum. The distribution of  distances is
not centered at zero, but
centered around $10^{-5}$~nm, due 
to numerical uncertainties in the exact location of the 
local minimum during the minimization procedure.
For further information on the procedure, and for a critical
discussion of the results obtained with this procedure, see
Ref.\protect\cite{lanavelong}.
\end{references}
\end{document}